\documentstyle[aps,preprint]{revtex}
 \tightenlines
\begin{document}
\draft
\title{Self-organized Critical Model Of Biological Evolution}
\author{H. F. Chau$^1$\footnote{Present Address: School of Natural Sciences,
 Institute for Advanced Study, Olden Lane, Princeton, NJ 08540.}, Louis
 Mak$^2$, and Peter K. Kwok$^3$}
\address{
 $~^1$Department of Physics, University of Illinois at Urbana-Champaign,\\
 1110 West Green Street, Urbana, IL 61801-3080, U.S.A.\\
 $~^2$Coordinated Science Laboratory, University of Illinois at
 Urbana-Champaign,\\ 1308 West Main Street, Urbana, IL 61801-2307, U.S.A.\\
 $~^3$Department of Mathematics, University of Illinois at Urbana-Champaign,\\
 1409 West Green Street, Urbana, IL 61801-2975, U.S.A.
}
\date{\today}
\preprint{}
\maketitle
\mediumtext
\begin{abstract}
 A punctuated equilibrium model of biological evolution with relative fitness
 between different species being the fundamental driving force of evolution is
 introduced. Mutation is modeled as a fitness updating cellular automaton
 process where the change in fitness after mutation follows a Gaussian
 distribution with mean $x>0$ and standard deviation $\sigma$. Scaling
behaviors
 are observed in our numerical simulation, indicating that the model is
 self-organized critical. Besides, the numerical experiment suggests that
models
 with different $x$ and $\sigma$ belong to the same universality class.
\end{abstract}
\medskip
\narrowtext
 Biological evolution takes place in bursts separated by relatively long
periods
 of quiescence \cite{Gould,Punctuated1,Punctuated2}. In fact, extinction may be
 episodic at all scales \cite{Raup}. To capture the essence of this scaling
 behavior, a punctuated equilibrium model of biological evolution is introduced
 recently by Bak and Sneppen (BS Model) \cite{Bak1}. An ecosystem in the model
is
 made up of $N$ species, each associated with a scalar between zero and one,
 called fitness of the species. The higher the fitness, the more adaptive to
the
 ecosystem and hence the less likely to mutate the species is. Thus, the
species
 with the minimum fitness in the system is most likely to mutate or extinct
next.
 Fitness is therefore also a measure of the barrier against mutation.
\par
 In the one-dimensional BS model, the $N$ species are arranged on a line with
 periodic boundary condition. At each timestep, the species with minimum
fitness
 and its two nearest neighbors mutate; their fitnesses are replaced by uniform
 and uncorrelated random numbers between zero and one \cite{Bak1}. The updating
 process is repeated forever. Scaling behavior is observed in the distribution
of
 distance between successive mutations once the stationary state is reached,
 which is a signature of self-organized criticality. Generalization to the
 $N$-dimensional and the mean field models can be made, and the results are
 essentially the same as their one-dimensional counterpart \cite{Bak2,Bak3}.
\par
 However, there are a number of conceptual difficulties in the BS model. First,
 it is unclear why mutation changes the fitness of the neighboring species and
 why such changes take place all at the same time when the least-fit species
 mutates. This kind of ``action at a distance'' couplings between the least-fit
 species and its neighbors are not physically justified. Second, there is no
 correlation between the fitnesses of a species before and after mutation (or
 extinction). This is not consistent with the fundamental assumption of the
 Darwinian school that all existing species share a common ancestor
 \cite{Evolution}. Finally, once the stationary state is reached, the average
 fitness of the species becomes a constant. Thus species in the present time
and
 those in the past have the same average adaptive power to the environment,
which
 is not true in nature.
\par
 Now we are going to introduce an alternative model of biological evolution
free
 from all the above conceptual difficulties. Our ecosystem is made up of a
single
 food chain consisting of $N$ species labeled from 1 to $N$, with 1 being the
 producer, and $N$ being the terminal consumer. The fitness of the $i$-th
species
 is denoted by $f_i$. Similar to the case of energy in physics, the value of
 fitness is meaningful only when it is measured with respect to a reference
 fitness level. The actual fitness value of a species is not a physical
 observable because the value of the reference fitness can be arbitrarily
 assigned. The relative fitnesses between different species (together with the
 environment) are the only physical measurable and meaningful quantities when
 describing how much adaptive power the species have.
\par
 In general, the ``relative fitness'' of a species can be written as
\begin{equation}
 v_i = \beta_i ( f_i - f_E ) + \sum_{k\neq i} \alpha_k (f_i - f_k)
 \label{E:Basic}
\end{equation}
\noindent
 for some $\alpha_k, \beta_k \geq 0$ where $f_E$ is the ``fitness'' of the
 environment. The first and second terms in Eq.~(\ref{E:Basic}) represent the
 competition of the species with the environment and with the other species
 respectively. In our present model, only the special case where $v_i = 2f_i -
 f_{i+1} - f_{i-1}$ is considered. Our ``nearest neighbor interaction
 assumption'' is justified provided that competition is only important between
 the predator and the prey. Further analysis in the case where the
environmental
 factors cannot be neglected is underway \cite{Further}.
\par
 Relative fitness can be also regarded as a measure of the barrier against
 mutation \cite{Bak1}. Provided that the probability of mutation of a species
in
 unit time is history independent, the characteristic mutation time for this
 species is given by:
\begin{equation}
 g(v) = Ae^{\lambda v} \label{E:Exp}
\end{equation}
 for some $A,\lambda > 0$. The above interpretation is consistent with the NK
 Model and its variations \cite{NKModel} in which mutation is regarded as
hopping
 between nearby locally optimal ``fitness landscape''. The characteristic time
 required for mutation therefore depends exponentially on the activation
barrier.
 Given the fact that some species mutate in a much faster time scale than
others,
 we expect $\lambda \gg 0$ \cite{Bak1,Chau}. With a sufficiently large
$\lambda$,
 the species with the smallest relative fitness is almost surely to be the next
 one to mutate or extinct. A lower bound for $\lambda$ will be given later in
the
 text.
\par
 With this assumption in mind, the species $i$ with minimum relative fitness is
 always chosen to mutate by updating its fitness $f_i$ as:
\begin{equation}
 \left( f_i \right)_{new} = \left( f_i \right)_{old} + B_{x,\sigma}
 \label{E:Update}
\end{equation}
 where $B_{x,\sigma}$ is a random number chosen from a Gaussian distribution
with
 mean $x$ and standard deviation $\sigma$. We choose $x > 0$ to reflect the
trend
 that newly mutated species is likely to be more adaptable to the ecosystem.
 Moreover, we require $\sigma \agt x$ so as to allow exceptions to the general
 trend. The relative fitnesses of all species in the ecosystem are then
 recalculated. Clearly, only the relative fitnesses of species $i-1$, $i$, and
 $i+1$ will change after the mutation of $i$. A single cellular automaton
 timestep is now completed, and the process is repeated indefinitely. The above
 construction is possible if we assume that the mutation (or replacement) of a
 species in our ecosystem does not alter the structure of the food chain.
Unlike
 the BS model, correlation between new and old fitnesses of the mutating
species
 is properly accounted for. In our model, the change in evolutionary landscape
of
 a neighboring species $j$ is a result of the coupling between the mutating
 species $i$ and its neighbor $j$ via the change in relative fitness $v_j$. The
 absolute fitness $f_j$ of the species remains unchanged right at the time of
the
 mutation of $i$. In this way, relative fitnesses may be regarded as ``fields''
 which couple between neighboring species, telling them the status of their
 neighbors. Our ``particle-field-particle'' interaction model looks more
natural
 than the ``action at a distance'' BS model.
\par
 We complete the model using the closed boundary conditions, namely, $v_1 = f_1
 - f_2$ and $v_N = f_N - f_{N-1}$. Closed boundary conditions are consistent
with
 the fact that $f_i$ are not physical observables: once we know the values of
 $v_i$, $f_i$ can only be determined up to a common additive constant. Finally,
 we would like to emphasize that the real physical time elapsed in each
cellular
 automaton timestep is not a constant and depends on the minimum relative
fitness
 of the system \cite{Bak1,Chau}.
\par
 By working with the relative fitness instead of the absolute one,
 Eq.~(\ref{E:Update}) can be rewritten as
\begin{equation}
 \left\{ \begin{array}{lcl} \left( v_{i-1} \right)_{new} & = & \left( v_{i-1}
 \right)_{old} - B_{x,\sigma} \\ \\ \left( v_i \right)_{new} & = & \left( v_i
 \right)_{old} + 2B_{x,\sigma} \\ \\ \left( v_{i+1} \right)_{new} & = &
 \left( v_{i+1} \right)_{old} - B_{x,\sigma} \end{array} \right.
 \label{E:BUpdate}
\end{equation}
 except possibly at the system boundaries. Eq.~(\ref{E:BUpdate}) is similar to
 the updating rules used by Bak and Sneppen in \cite{Bak1}, with the important
 exception that the random variables we used for all the three species are
 correlated. We therefore expect our model to exhibit self-organized critical
 phenomena because it is observed in the ``less correlated'' BS model. This is
 indeed the case as we proceed to show it numerically.
\par
 Numerical simulation is carried out with $N = 1024$. We fix $x$ to be 1.0, and
 systems with other values of $x$ can be rescaled to ours by a simple scale
 transformation. We have considered cases with $\sigma$ between 0.1 and 3.0. In
 all cases, after a long transient period, the system settles into a stationary
 state independent of the initial fitnesses of the species. Then a further
 $2\times 10^8$ mutations are recorded for statistical purpose. The
distribution of
 the relative fitness $v$ and the minimum relative fitness $v_{min}$ are shown
in
 Fig.~\ref{F:RELFIT} and Fig.~\ref{F:MINFIT} respectively. Similar to the BS
 model, there is an upper critical value $v_c$ above which there is zero
 probability of the minimum relative fitness $v_{min}$. Although the precise
 values of $v_c$ (see Table~\ref{T:C_V}) is $\sigma$ dependent, the shapes of
the
 distribution curves do not change. Unlike the BS model, the distribution of
$v$
 (and $v_{min}$) does not flatten when its value is greater than (or less than)
 $v_c$. In fact, Fig.~\ref{F:MINFIT}b tells us that (except for the small hump
 near $v_c$) the distribution of $v_{min}$ ($D(v_{min})$) can be well
 approximated by
\begin{equation}
 D (v_{min}) \approx \left\{ \begin{array}{ll} B \exp \left( \nu v_{min}
\right)
 & \hspace{0.2in} \mbox{if~} v_{min} < v_c \\ 0 & \hspace{0.2in}
\mbox{otherwise}
 \end{array} \right. \label{E:Dv_min}
\end{equation}
 for some $B$, $\nu$ $>$ 0. From Eqs.~(\ref{E:Exp}) and~(\ref{E:Dv_min}), the
 distribution of time between successive mutations ${\cal D} (T)$ is then given
 by \cite{Chau}:
\begin{equation}
 {\cal D} (T) = \frac{1}{\lambda T} D (\frac{1}{\lambda} \ln \frac{T}{A})
\approx
 \frac{B}{\lambda} \left( \frac{1}{A} \right)^{\frac{\nu}{\lambda}}
 T^{\frac{\nu}{\lambda}-1} \label{E:DT_Scaling}
\end{equation}
 whenever $T \leq Ae^{\lambda v_c}$.
\par
 The average mutation rate of a species is $\int D(v)/g(v) dv$, and the minimum
 possible mutation rate of the least fit species is $1/g(v_c)$. The assumption
of
 always choosing the least adaptable species to mutate is justified provided
that
 the sum of mutation rates of all the other species is much less than the
 mutation rate of the least adaptable one \cite{Chau}; thus $(N-1) \int
D(v)/g(v)
 dv \ll 1/g(v_c)$. This is true if $\lambda \agt N$. So for the case where $N$
is
 sufficiently large, $\nu /\lambda$ is negligible and hence $1/T$ scaling is
 expected for ${\cal D} (T)$.
\par
 Suppose $i$ and $j$ are the mutating species at the present and the previous
 timesteps respectively. We define the distance from the last event $d$ as
 $|i-j|$. The distribution of $d$ ($F (d)$) is shown in Fig.~\ref{F:CORRDIS}
 where $1/d^\alpha$ scaling is observed for about 2.5 decades with $\alpha =
2.87
 \pm 0.02$, which is independent of the values of $x$ and $\sigma$ used in all
of
 the simulations. So we strongly suspect that the critical exponent $\alpha$ is
 independent of $x$ and $\sigma$, and models with different $x$ and $\sigma$ in
 fact belong to the same universality class. Furthermore, scaling behavior in
 both ${\cal D} (T)$ and $F (d)$ are clear signs of self-organized criticality,
 that is, under its own dynamics, the ecosystem is driven into a state where
 there is no characteristic correlation length and time scales.
\par
 For a given value of relative fitness $v_{ref}$, an avalanche of size $s$ is
 defined to be the process of $s$ consecutive mutations with minimum relative
 fitnesses less than or equal to $v_{ref}$, which are immediately preceded and
 followed by mutations with minimum relative fitnesses greater than $v_{ref}$
 \cite{Bak1}. For sufficiently large $\lambda$, the distribution of avalanche
 size coincides with that of the total number of mutations occurred in a time
 interval of $g(v_{ref})$ \cite{Chau}. Fig.~\ref{F:ACTIVITY} shows that
 $1/s^\beta$ scaling holds over three decades; the value of the critical
exponent
 $\beta$ is found to be 0.99$\pm$0.02 and independent of the value of $\sigma$.
\par
 In summary, we have introduced a self-organized critical model of biological
 evolution in which competition between between species is the driving force
 using the idea of relative fitness. Relative fitness acts like a background
 field and tells the species to response to its own situation relative to the
 other species and to the environment. Although the average value of relative
 fitness is a constant in the system, the average absolute fitness increases
with
 time, reflecting the improvement of adaptive ability. The time-average
relative
 fitness of a species is constant which is consistent with the Red-Queen
 hypothesis \cite{RedQueen}. Our numerical experiments suggest that models with
 different values of $x$ and $\sigma$ belong to the same universality class.
Our
 model may also belongs to the same universality class as the Robin Hood model
 whose critical exponent $\alpha = 2.86\pm0.01$ \cite{Robin}. Numerical
 experiments using periodic boundary conditions ($v_1 = 2f_1 - f_2 - f_N$, $v_N
 = 2f_N - f_1 - f_{N-1}$), and reflective boundary conditions ($v_1 = 2f_1 -
 2f_2$, $v_N = 2f_N - 2f_{N-1}$) are also performed. We found that the shape of
 all the distribution curves, together with the critical exponents, are
 insensitive to the boundary conditions used.
\acknowledgments{Interesting discussions with Henry K. Kwok, K. Sneppen, and T.
 Hwa are gracefully acknowledged. This work is supported in part by the NSF
grant
 PHY-9100283 and DOE grant DE-FG02-90ER40542.}

\begin{table}
 \caption{Upper critical relative fitness $v_c$ as a function of $\sigma$. In
all
  the cases listed below, $N$ = 1024 and $x$ = 1.0.
  \label{T:C_V}}
 \vspace{0.2in}
 \begin{tabular}{ll}
  $\sigma$ & $v_c$ \\ \hline 0.5 & $-0.44\pm$0.01 \\ 1.5 & $-0.72\pm$0.01 \\
3.0
  & $-1.23\pm$0.01
 \end{tabular}
\end{table}
\begin{figure}
 \caption{Distribution of relative fitness $v$. $\sigma$ = 0.5, 1.5, and 3.0
are
  used in the dotted, dash, and solid curves respectively. This convention is
  used in all subsequent figures. \label{F:RELFIT}}
\end{figure}
\begin{figure}
 \caption{Distribution of minimum relative fitness $v_{min}$. This is shown in
  linear plot in (a) and in semilog plot in (b). \label{F:MINFIT}}
\end{figure}
\begin{figure}
 \caption{Distribution of the distance from last event $d$. Power law behavior
  is observed over 2.5 decades in all the three cases, and their critical
  exponents agree with each other and are equal to 2.87$\pm$0.02.
  \label{F:CORRDIS}}
\end{figure}
\begin{figure}
 \caption{Distribution of avalanche size below a reference fitness level
  $v_{ref}$, which is chosen in the way that about 98\% of the minimum
fitnesses
  will fall below it. These values correspond to $-1.4$, $-0.8$ and $-0.47$ for
  the solid, dash, and dotted curves respectively. Scaling behavior is observed
  over 3 decades in all the three curves. The scaling region will lengthen as
  $v_{ref}$ approaches $v_c$. The critical exponent $\beta$ is measured to be
  0.99$\pm$0.02 in the regions where the power law holds. \label{F:ACTIVITY}}
\end{figure}
\end{document}